# NETWORK WEAVING TO FOSTER RESILIENCE AND SUSTAINABILITY IN ICT4D

Mario A. Marais, CSIR, South Africa, mmarais@csir.co.za

Sara Vannini, Information School, University of Sheffield, United Kingdom, s.vannini@sheffield.ac.uk

**Abstract:** A number of studies in Information and Communication Technologies for Development (ICT4D) focus on projects' sustainability and resilience. Over the years, scholars have identified many elements to enable achievement of these goals. Nevertheless, barriers to achieving them are still a common reality in the field. In this paper, we propose that special attention should be paid to communities' relationships, self-organizing, and social capital - and the people's networks that enable them - within ICT4D scholarship and practice, as a way to achieve sustainability and resilience. Building on Green's work (2016) on social change as a force that cannot be understood without focusing on systems and power, we claim that ICT4D would benefit from intentionally growing social capital and fostering networks within its systems. We propose "network weaving" (Holley, 2013) as a practical approach, and we explore its potential to complement and advance existing ICT4D frameworks and practices, including the sense of community of the researchers themselves**.**

## 1. INTRODUCTION

### 1.1 The scope of sustainability and resilience inquiry

Over the years, scholarship in Information and Digital Technologies for Development (ICT4D) has extensively engaged with issues of projects sustainability and resilience, defined jointly as the ability of systems to withstand, recover from, adapt to, transform, and endure amid external stressors and over time (Heeks & Ospina 2019). Both barriers to achieving them and ways forward for the field have been identified, generating theoretical and methodological frameworks.

In this research, we consider some of the ways sustainability and resilience have been addressed in the ICT4D literature, and we propose a way to complement existing frameworks by focusing on particular *fundamental values* applied to growth of social capital and fostering of networks. We maintain that the focus of research should shift from the - often temporary - project level to a more systemic level, which considers communities in all their contextual complexities and is designed to contribute to sustainable development processes. The next section, thus, deals with two scopes of inquiry, the project level and the community-in-context level.

## 2. THE SCOPE OF SUSTAINABILITY AND RESILIENCE INQUIRY

### 2.1. Project-level sustainability

Project-level sustainability is the context in which the different value systems of ICT4D key role players have an immediate and major influence on sustainability, and ultimately resilience. Despite the extensive research uncovering how to make ICT4D projects sustainable and resilient, there are still many obstacles to realize the desired change we want to bring, and many projects still fail (Fierbaugh, 2017; World Bank, 2011). Research on ICT4D failures is quite extensive, with a large focus on sustainability-connected issues. From "design-reality gaps" (Heeks, 2002) and the need to promote a stronger emphasis on local dynamics and socio-cultural factors, to the need to analyze power dynamics and the role of researchers (e.g. Avgerou & Walsham, 2000; Unwin, 2009; Tedre, Sutinen, Kähkönen, et al., 2013; Brunello, 2015), through employing participatory approaches to ensure projects are socially sustainable and empowering for communities (e.g. Gómez, 2013; Heeks, 2008, 2010), a considerable amount of work has also looked at viable approaches to successful projects.

For example, participatory methods have been used to promote community empowerment and ownership, thus increasing the probability of project sustainability (Bentley, Nemer, & Vannini, 2017; David et al., 2013; Vannini, Rega, Sala, et. al, 2015). Participatory work has also been promoted as a way to engage as equal partners and develop local capabilities and resilience. A number of studies guided by strong and explicit value systems identified that respecting local traditions, engaging in longer partnerships, co-designing at all levels, and promoting mutual ownership and shared decision-making, are key to achieve sustainable outcomes (Bidwell, 2020; Rey-Moreno, Roro, & Tucker, et al., 2013; Winschiers-Theophilus, Chivuno-Kuria, Kapuire, et al., 2010). The concept of participation *per se* has also been questioned. Researchers have been questioning what and whose voices and knowledges are eventually heard, and which ones are lost along the way (Halabi, Sabiescu, David, et al., 2015; Russ, 2021). The relationships between researchers and communities have also been examined in terms of power dynamics, and frameworks were offered to rethink mismatching expectations when dealing with intercultural contexts (Brunello, 2015; Vannini, Nemer, Halabi, et al., 2017). This shows progress towards respect and sensitivity towards differences in value systems.

Sustainability, as bottom-up alignment of communities' interests enabled by building social resources and social capital, has a rich tradition in development and ICT4D research. Marais (2016) provides a comprehensive overview of social capital in ICT4D. Renken and Heeks (2018) review the role of Social Network Analysis. Social capital was described in terms of networks of relationships based on trust by Farr (2004). Sein, Thapa, Hatakka, et al. (2019) propose three groups of theory underpinning the ICT, D, and '4' in ICT4D, including theories of Social Capital and Actor-Network theory (ANT). The importance given to the role of social capital here shows how social interactions enable action through the enhancement of alignment, trust, and acceptance. Trust is considered pivotal, as it is rooted in different value systems, and open and equal engagement is required for mutual trust. This is where the work of June Holley, which we will introduce in section 3, is relevant to the enduring problems in ICT4D.

The integrative concept of 'sustainable livelihood security' by the World Commission on Environment and Development (WCED) reflects the connections between basic needs, secure resource ownership, and long-term resource productivity (WCED, 1987). Chambers & Conway (1992) include social resources and capabilities (Sen, 1984) in their definition of sustainable livelihoods. Social capital became embedded in the systemic perspectives on sustainable development partly due to its prominence in the World Bank-based research development policies (Woolcock & Narayan, 2000) and the linkages between sustainable development approaches and the Choice Framework (CF) of Kleine (2010; see also Grunfeld, Pin & Hak. 2011).

The social capital of a community is grown by local individuals who can fulfil many important roles. Examples include: local champions who facilitate interaction with development initiatives (Renken & Heeks, 2019), intelligent intermediaries (Gopakumar, 2007), infomediaries (Mukerji, 2008), or social connectors (Díaz Andrade & Urquhart, 2010). These individuals are vital for success in ICT4D projects, particularly for promoting access to information and adopting and fostering innovation and use of social capital (Madon, 2007; Sey & Fellows, 2009; Gómez, Fawcett, & Turner, 2012; Marais, 2016).

However, complex challenges remain inherent to current development funding schemes. Recent research has underlined *the ethical dilemma* between designing for participation and respecting the constraints of funders, who usually budget for short-term projects that don't allow for the continuity needed in development. The derogatory phrase often used in the literature is that of "parachuting" into a community followed by a quick departure (Holeman & Barrett, 2017, p.920, Unwin, 2014, Raftree, 2011, 2018). Cleary, this is a particular type of value system in action.

### 2.2. Community-in-context sustainability

In a recent report addressing the drivers of change in human development, Knox Clark claims that decision-making in the sector is still mainly not guided by evidence, despite supporting data being provided. This is attributed either to decision-makers being affected by a number of other considerations - e.g., politics, resource availability, security - or to decision-making not considering evidence as an important component of their processes (Knox Clarke, 2017). For example, many donors seem to prefer funding project-based initiatives, while evidence shows the *greater value of core and long-lasting funding* to organizations. ICT4D itself is often operated in under-resourced systems where long-term projects are not the norm.

Green's work (2016) suggests that social change cannot be understood without focusing on systems and power. Social change progresses must address both how power is distributed and how it is possible to re-distribute it within and between social groups. Power and systems are always situated, affected by and affecting complex social contexts (Green, 2016). For example, ICT4D processes are often directed or funded by the North, and it is rare that local or South-South collaborations are enabled and amplified (Walsham, 2020). Funding for collaboration in ICT4D between the Global North and the Global South has been scoped by UN agencies such as USAID, or science foundations such as the UK ESCRC or the Swiss DEZA as creating shared ownership and equal partnerships. In practice, though, funded projects are often directed by the Global North expertise, while countries in the Global South are mainly labelled as beneficiaries. When South-leading collaborations are enacted, partners are sometimes not considered as equally contributing (van Stam, 2020), which reflects a judgment that may be based on a value system. Under these circumstances, project sustainability is hardly enabled. Short-term funds will often ignore the hidden running costs that maintaining systems long-term will require, and North-enabled projects will often present structural power dynamics that are hard to discern and overcome.

As mentioned in Section 2.1, individuals play many necessary roles, but these roles can be misused as well, or divide communities due to patronage issues. The reliance on individuals as infomediaries or local champions to ensure sustainability, resilience, and community ownership may also have unwanted consequences. First of all, local champions might not be easy to find in any given community. Secondly, they have characteristics such as being innovative, natural leaders, dynamic, strategic, propensity to take risks (Renken & Heeks, 2019) that could potentially lead them to leave their communities for better opportunities elsewhere. Thirdly, a too great emphasis on champions' responsibility reinforces the narrative of "the genius" extraordinaire. This might be problematic as it tends to over-burden the individual with the responsibility to create change instead of focusing on more systemic power imbalances. Also, the concept intrinsically bears individualistic values (of "the" champion) that are not necessarily the most appropriate for work with more collectivist non-Western cultures (Arora, 2019; Jimenez & Roberts, 2019; Oliveira, Muller, Andrade, et al., 2018). This illustrates a difference in value systems that prioritize the individual's actualization versus the loyalty towards being part of a community. A shift to a more collective idea of championship would be more fitting for the kind of work ICT4D is pursuing.

In this paper, we argue that more attention should be given not only to the family of approaches that follow bottom-up, participatory strategies in ICT4D, but also to possible differences in value systems and frameworks that leverage on communities' self-organising, and that build on existing relationships within and between communities, without relying on single individuals only. These existing relationships may already have (and usually already have) led to innovation, development, and social change in their communities. They might be more resilient and sustainable in time, as they are based on existing and enhanced social capital as enabled by shared value systems.

We propose a network approach that builds on existing experiences and scales up community level development already led by a number of actors, including community champions and people less on the spotlight, as well as Non-Governmental Organizations (NGOs), Faith-Based Organizations (FBOs) and Community-Based Organizations (CBO)s. Intentionally enabling networks within the systems in which ICT4D projects operate requires adopting a relationship and social capital framework as one of the foundations for sustainable development. This could be done in the context of periodic external interventions (e.g. projects relying on external funding for development frameworks), which contribute to long-term, sustainable development, but are hardly the main drivers for it. We propose the practice of "network weaving" (Holley, 2013) and the value system on which it is built, and we explore its potential to complement and advance existing ICT4D frameworks.

## 3. NETWORK WEAVING AND ITS POTENTIAL FOR ICT4D

In this section, we talk about what network weaving is and how it addresses change through situated, transformational responses to systemic power structures. Finally, we identify how this approach can bring forward ICT4D research.

### 3.1. What is Network Weaving?

The concept and practice of "network weaving" was started by June Holley (2013) to help low-income entrepreneurs in the Appalachian region in Ohio, one of the poorest regions of the U.S., create networks that would bring opportunities and change to their communities.

According to Holley (2013), networks are complex sets of relationships among people and communities that have the ability to address power imbalances by 1) encouraging intentional peer relationships which recognize the value and contribution that individuals can make; 2) considering every individual to be a leader, who has the ability to connect and initiate collaborations. In this way, "power is distributed, not concentrated" (ibid., p.10); and 3) including all stakeholders' voices in the process of generating change. This is a summary of the elements of this value system.

Networks are at the basis of societal systems. The status quo is held in place by old networks, and exposing them, according to Holley, is paramount to change. Societies and communities also consist of unconnected or loosely connected people. Holley claims that, to shift systems and rebalance power, it is important for people to understand the old networks and subsequently foster new networks that can reposition peoples that have been historically relegated at the margins of networks to their centre. This re-positioning of and within networks, together with the attention to enable grassroot networks to emerge and thrive, can effectively transform systems and enable change (ibid.).

Networks can be of four different kinds, at time co-existing, but especially interlocking and complementing one another so that a network can function (see Fig. 1). First, they have to be intentional, meaning they have to have the same goal or share the same vision. These networks can be formally organized or not. Second, they have to establish relationships. Intentional and relationship networks are connected to one another. Holley (ibid.) identifies the lack of attention to relationships as the main cause for network failures, as it can lead to issues such as lack of trust or lack of support. As argued previously by the authors, trust involves in-depth engagement with value systems. Third, they have to be action-oriented and self-organizing. Fourth, they have to have systems in place to provide support and foster accountability for their work.

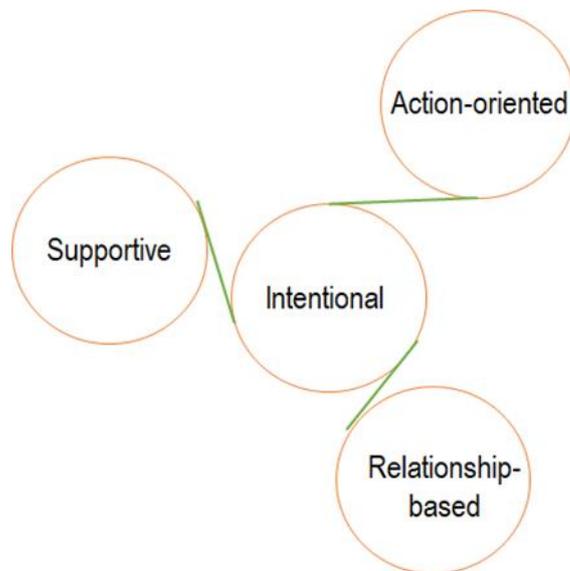

**Figure 1. The four Interlocking Networks, adapted from Holley, 2013**

Network weaving indicates being intentional about networks and fostering them. This can be done by "connecting people strategically where there's potential for mutual benefit, and serving as a catalyst for self-organizing groups" (ibid., p.24). Figure 2 summarizes the four different roles Network Weavers can have as identified by Holley: two are concerned with fostering the network (Connector Catalyst and Guardian), and two are action-oriented (Self-organized Project Coordinator and Facilitator). Also, Connector Catalyst and Project Coordinator tend to operate on the micro-level, e.g. strengthening ties that are only loosely connected or seeing opportunities for a network, while Network Guardian and Facilitator tend to work on the macro-level, e.g. improving networks' systems or helping people see the benefits of a network approach.

|  | micro-level | macro-level |
|---|---|---|
| **build** | **Connector Catalyst** Connects people, gets network building started | **Guardian** Helps put in place all systems needed for networks: communication, training, support, resources |
| **act** | **Self-organized Project Coordinator** Helps co-ordinate self-organized projects | **Facilitator** Helps convene people to set-up a more explicit and focused network |

**Figure 2. Network Weaver Roles, adapted from Holley, 2013.**

In network weaving, leadership is distributed and everyone can be a leader (or, in ICT4D terms, a champion) because everyone is believed to manifest leadership abilities. These abilities can also be learned, improved, and leveraged upon. The self-organizing nature of network weaving is what encourages people to take initiative, ultimately fostering ownership and leading to transformation. As everyone is a leader, the success of a network depends on each and every participant into the network, and on how leaders are supported.

## 3.2. Network Weaving's value for implementing sustainability and resilience in ICT4D

Several characteristics of network weaving make it a suitable addition to existing frameworks addressing ICT4D projects sustainability and resilience. In this section, we underline the existing linkages between them and ICT4D theorized best practices.

First, network weaving focuses on **co-creation** processes and on all stakeholders' **participation**. Networks include diverse stakeholders, with their different approaches and perspectives. Stakeholders are all nodes of the networks, organized in a non-hierarchical way. A network mindset implies decentralized decision-making and collective action - i.e. it would require that local communities participate in development strategy decisions that are usually made by Government, NGOs and International Development Organizations. Living Labs (LLs) are a similar development strategy, which describes the Finnish Triple Helix models of collaboration between industry, government and citizens (Følstad, 2008), where product development shifts from the laboratory to the real-life context of citizens (the "living lab") at home and at work, where co-creation of innovations can emerge. In ICT4D Finnish development aid has assisted the development of ICT-enabled rural LLs in South Africa (SA) (Stillman, Herselman, Marais, et. al, 2012). A network mindset requires contextual development strategies where the diversity of interests is represented, and the space is held for multiple ways of achieving outcomes, in which indigenous knowledges, various value systems and so-called expert knowledge can interact and evolve. The approach also responds to the need of developing adaptive capabilities, which can innovate and respond to the complexity of our planet transformations.

Second, network weaving builds on existing literature on the importance of intermediaries, infomediaries and champions. By saying that everybody can be a network weaver, the focus on **leadership** is shifted to a shared, collective idea of "championship", which is more inclusive, more attuned to collective dynamics and value systems that are usually more common in non-Western communities, and possibly more sustainable in the long term. This different idea of leadership is especially relevant when it comes to focusing on women's participation in development leadership. Women as leaders and champions are often hidden. For example, schools in SA depend largely on women teachers, but School Principals are mostly male (Skosana, 2018). Activities related to care work are predominantly performed by women and are often devalued and made invisible in the information field, including in spaces such as community centers and access to information venues where this kind of caring work is paramount (Sweeney and Rhinesmith, 2017). Despite the centrality of their work, including their reproductive labour, women are hardly recognized in society capitalist production (Federici, 2012; Stillman, Sarrica, Anwar, et al., 2020). Also, women in resource-constrained contexts form very strong relationships, usually out of necessity, and they are more likely to use their social capital to survive because of their position in society (Stillman et al., 2020). Women's positions and positionality, then, makes them natural networks weavers. A focus on networks is relevant because it recognizes different ways of leading and it might help emphasise women's leadership and the values that are often associated with women's roles.

Third, this approach is **empowering** for communities, not only as they have more access to decision-making and to leading the change they want to see, but also because it enables them to be aware of the networks they are embedded in and of the power they have to shape decisions. This **awareness** also brings with it more exposure to a plurality of ideas, and more peer learning.

Finally, the approach speaks to ICT4D scholarship's focus on **transparency** as a way to achieve social justice (Smith, 2014). A network approach encourages sharing by default, including not only sharing outcomes and approaches from different projects, but also sharing networks themselves, so they can grow and evolve. Networks are intrinsically flexible: they often have no clear boundaries, which ensures they have a great potential for expansion. This natural evolution of networks is beneficial to ensure best practices' **dissemination**: while projects can come and go, and academic findings rarely reach communities outside of researchers' own influence sphere, networks learning exchanges at the community level can self-sustain as new members become involved.

In each of these categories, trust relationships are the foundation of community development and social change. This is in line with Toyama's view (2010) that "technology—no matter how well designed—is only a magnifier of human intent and capacity."

The network weaving approach that we are hereby proposing is the distillation of many years of experience in communities and implementation of a value system (Holley, 2013). Holley's work provides practical advice and clear guidelines on how to operationalize the practices at the micro and macro levels, including how to recognize, support, and deploy networks into a project. Also, this approach fits in with the way that people are used to develop and assist each other in informal and formal ways, especially within the kinds of resource-constraint contexts where ICT4D operates.

## 4. APPLICATIONS IN CONTEXT

This section addresses how aspects of network weaving have already been applied in the ICT4D sector. Two examples are provided to show benefits and challenges of the approach as connected to Holley's work.

### 4.1. Rural entrepreneurs in Southern Africa

Twenty years of experience of the challenges in ICT4D and "technology for development" implementations in Southern Africa is distilled in a paper by Van Rensburg, du Buisson, Cronje, et. al. (2014). Their challenge is achieving a direct linkage between ICT4D application and adoption to the process of scalable socio-economic development, which are elements that directly connect to sustainability and resilience. The largest resource for local development remains the local, regional (provincial), and the national government. The autonomous nature of these three spheres of government is highly dependent on excellent collaboration, which is problematic. This leads to a top-down development approach, where the richest consultations are at local government level, as mandated by law. In reaction to the lack of real collaboration between community and government, the authors create a development perspective that is founded in participatory processes during which both the individual and communal asset-base is improved, while making it part of a bottom-up process of developing a shared vision. This is integral to this value system. The major shift in focus of rural enterprise and economic development (REED), when compared to ICT4D, is the focus on *people-centred network development*. This focus is rooted in a rethinking of African development with a respect for the traditional African context and culture and the "African communal relationships of mutual responsibility that was once finely tuned to respond to the needs of the African community, climate and its environment" (Sparks 1990 as cited in Van Rensburg, et. al., 2014, p.2). The respect for African value systems is clear.

The people-centred network development corresponds to the *intentional kind* of network described in network weaving (see Fig. 1), in which people have a shared focus, or vision, and the basis is *relationship-based*, such as established communal relationships and the development of new relationship networks. Infopreneur® networks, as discussed below, correspond to the *action-oriented* kind of network that enables self-organizing. The support and accountability systems that complete the four inter-locking and complementary networks described by Holley are used as a value within the African communal relationships of mutual responsibility. These relationships are currently mostly augmented by the more formal structures, processes and relationships of development practitioners that have relationships with communities, governments and funding sources.

Infopreneur® networks play a more specific role than the other types of actors described previously, namely: intermediaries, infomediaries, or social connectors. Their focus is to improve the reach and nature of the extension services as normally delivered by government departments, in order to "support new scalable and sustainable micro-enterprises within the local contexts (ibid, p.1)." They interact directly with the community via "knowledge economy services" to foster the five main community assets: "human, physical, financial, natural and social." (ibid.).

A typical example is the development of a geo-located database of small businesses for a municipality that needs to know the businesses that are to be served by their Local Economic Development (LED) plan executed by their LED office. This plan gets funding via the Provincial government, which requires good data to inform plans, budgets and to execute monitoring and evaluation. Rural (agricultural) schools are also brought into the network to assist in the revitalisation of neglected irrigation schemes and small-holder agricultural activities (ibid.). And interesting point for our study is that Infopreneur® networks are seen as "a network of change agents which are micro enterprises within their communal contexts" (ibid., p.5). Hence, a "micro-franchise or micro-license" model was developed so that everyone could run their own businesses while aligning to the Infopreneur® network goals. Therefore, the basic approach is "development through enterprise" (ibid.), in which participants contribute to the network and also benefit from their membership. it. This is a large element in the REED values system.

The Infopreneur® networks can be mapped to the four different roles of Network Weavers (Holley, 2013). For example, via the building of community assets, the network combines elements of the Connector Catalyst role, by growing social assets, and the Self-organized Project Coordinator role, by coordinating the knowledge economy services. These two roles are mostly at the micro-level scale of communities in fostering networks and coordinating project level actions.

The Infopreneur® networks are supported by a Regional Infopreneur®, a business minded person with good networking and change management skills, who plays a vital role to connect and create bridges between communities' agendas, Infopreneurs®, and researchers (Van Rensburg, et. al., 2014). The Regional Infopreneur® is connected to a formal organisation providing the support systems for these networks and helping Infopreneurs® to build their own network to be more explicit and focused.

The Regional Infopreneur® provides mentoring and support to the Infopreneurs® and assists them *to internalise the values of the network* via the use of change management tools and processes. The value of being part of a mutually supportive network is vital in the REED values system. The change in ingrained mindsets (with the associated value systems) regarding community development is the biggest challenge they face, as they run against competing interests. For example, local officials are driven by quick turnarounds of outputs, to avoid having their yearly budgets cut by Provincial and National Treasury. This leads to quick-win projects with rapid delivery of tangible outputs, which are prone to corruption and against the principles of collaboration (versus competition) and the time required when building support networks (Marais, 2016).

This brief summary shows how the REED approach, with its focus on mapping Small, Medium and Micro Enterprise development and networks, does share commonalities with network weaving and its focus on community change, value systems and sustainable development.

In the next section we discuss the discovery of local support networks in an ICT for Education project.

### 4.2. Education in rural South Africa

SA has one of the most unequal school systems in the world with the widest gap between the test scores of the top 20% of schools and the rest (Amnesty International, 2020). The Council of Science and Industrial Research (CSIR) in Pretoria, SA, executed a three-year project from the Department of Rural Development and Land Reform (DRDLR) for implementing the DRDLR ICT for Education Project (ICT4E) (Herselman, Botha, Dlamini, et al., 2019). Teachers at 24 schools in seven of the nine provinces in SA were trained over a period of a year and a half to integrate mobile technology to support their teaching and learning. The aim was to integrate mobile technology into schools in order to facilitate improvement in the quality of teaching and learning. Thus, the scope of the project included a focus on providing Teacher Professional Development through accredited training materials (Botha & Herselman, 2018).

The focus of the project was resource-constrained rural schools characterized by lack of internet, remote geographical locations, unreliable electricity, low digital literacy and limited local funding sources, e.g., parents. (Botha and Herselman, 2016). During a project training session, the attendees were asked to tell us who are the types of people in their town that assists them with problems regarding devices, e.g., cell-phone, tablet or computer, or services such as emailing, internet searches and photocopying. The results are presented in Figure 3.

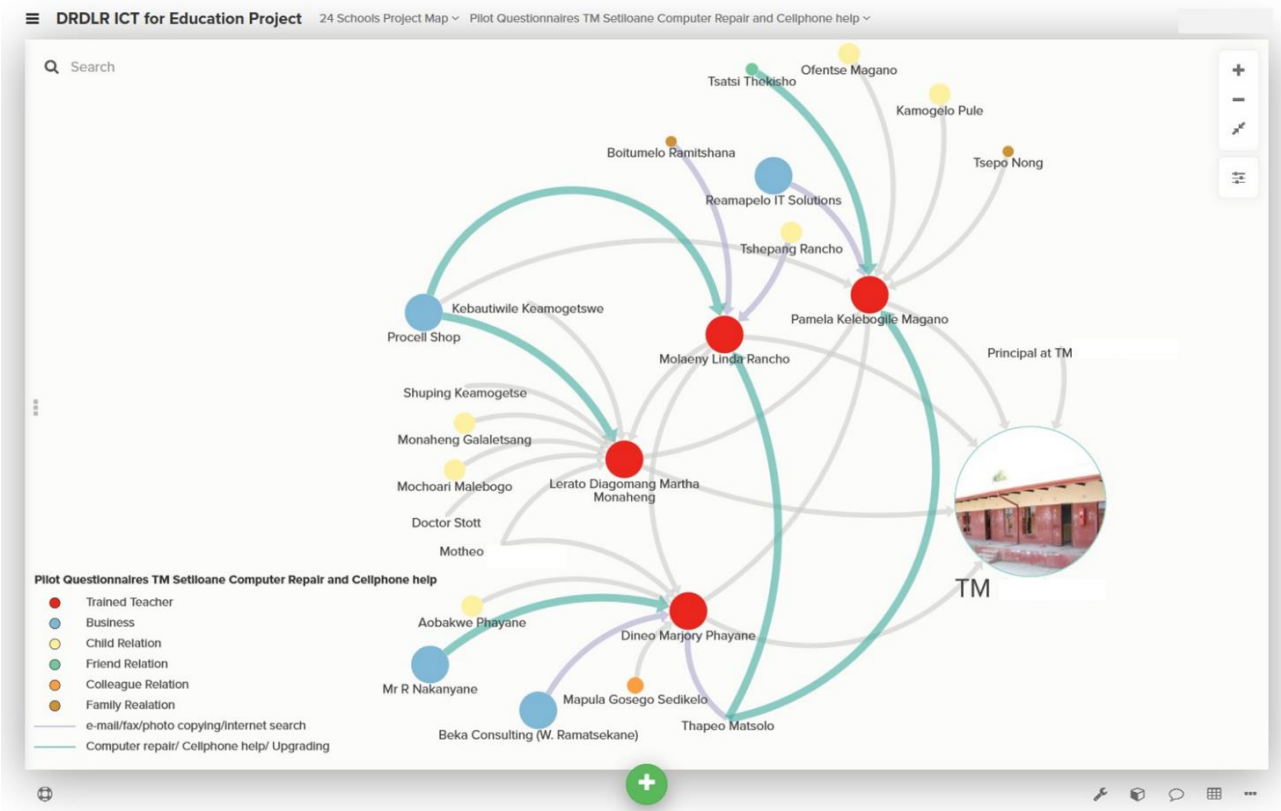

**Figure 3. The different types of relationships of school teachers at an ICT4E school (Marais, 2019).**

As the legend of the relationship map done in https://kumu.io shows, the four teachers of a school are represented in red circles. Businesses are shown in light blue, children in yellow, friends in green, family members in brown, and colleagues in orange. Technical support service relationships for devices are shown green and service relationships in grey. Two of the teachers had nine relationships, two teachers had eight and the remaining teacher had six relationships. Relationships with their school were not counted.

In Lerato's case, her nine relationships consisted of one relationship with a business, one with a Technical and Vocational Education and Training (TVET) college, two with teacher colleagues, two with her children, and three with friends. The direction of the arrows indicate that she provided support services in seven relationships: a relationship with one of her teacher colleagues, her children, three friends and the TVET college. In turn, she received support from another teacher colleague and a business.

The map indicates to teachers that they have many existing support relationships, including important peer-to-peer relationships. This value of helping each other is a key element for sustainability. Four technical support sources exist of which two are businesses and two were friends. Technical support services are thus balanced between businesses and social relationships with friends. Three of the teachers had technical support from a shared friend, which indicates the many shared relationships. This can be seen as the initial phase of network weaving where local champions are identified.

### 4.3. Discussion and Way Forward

In this paper, we presented a brief reflection on sustainability and resilience frameworks and discussion in ICT4D scholarship. Among others, we emphasized the value of approaches that focus on social capital and the way in which differing value systems are engaged with. We introduce network weaving as a possible fruitful addition to our scholarship's current efforts focusing on a strong value system of each person being a leader.

In our analysis, we show the characteristics of network weaving that makes it not only suitable for the field, but also a valuable addition to the frameworks that currently inform ICT4D's best practices. A network approach is participatory, co-creative, and empowering; it focuses on awareness raising, distributed leadership, and transparency. These characteristics speak to bringing change within and beyond single projects, by addressing the power structures that often prevent change from happening, as theorized by Green (2016) and the underlying value systems emphasised in this work.

Furthermore, network weaving is transformational. It is not focused on individual successes and heroic efforts, but on the possibility for everyone to be a leader responsible for the change they want to see. This resonates with ICT4D's aspirations to find more collective-oriented frameworks, especially when working in the Global South, as well as to the need to highlight the work that women often perform in these contexts (Sweeney and Rhinesmith, 2017). Also, a focus on networks underlines the importance of social capital, values, and relationships in ICT4D.

We argue that a focus on enabling and weaving networks should be included as an integral part of ICT4D thinking and as a component of ICT4D's projects. Building networks that can operate to build resources and capabilities, that can lobby for change, where people trust each other, are supported, and that can be sustained in the long-term will initially require intentional project design from day one.

We invite the broader ICT4D community to join us in representing ourselves as a network in ourselves, which may be used to add value to the attractiveness of the domain, by illustrating the diversity of the research agendas, disciplines and existing partnerships. In this kind of multidisciplinary research domain, it is difficult for newcomers, students, and local partners to get a grasp of what the reach and scope of this domain entails, and network relationship mapping will enable us to ease the introduction of new entrants and for existing members to develop new partnership opportunities. In addition, researchers have referred to the "significant amount of reinvention of the wheel in ICT4D" (Zheng, Hatakka, Sahay, et al., 2018, p.9) that may be reduced, as new entrants to the field tend to neglect earlier research since technologies change rapidly. We are convinced that the spirit of network weaving and the value base of a distributed leadership can enhance our relationships and transform our values, often still focused on individualistic Western perspectives, into more collective-oriented perspectives reflecting the diversity of challenges in our increasingly complex world.

While we believe network thinking and its underlying value systems provide a valuable addition to existing ICT4D's scholarship, we are aware of the limitations of our work and of the need to further explore and validate this approach. We have based our analysis on an exploration of past work in the field, where different characteristics of network weaving have emerged, but were not necessarily applied systematically. At the same time, we are aware that network weaving was first created within a Western context and, although it was developed and applied exactly to empower resource-constrained communities and redistribute power, communities in the Global South might receive it differently. We encourage more collective research to embrace this approach and document it in ICT4D, both in Global North and Global South contexts.